\documentstyle[prb,multicol,aps,epsf]{revtex}
\newcommand	\be	{\begin{equation}}
\newcommand	\bea {\begin{eqnarray}}
\newcommand	\ee	{\end{equation}}
\newcommand	\eea {\end{eqnarray}}

\newcommand	\bi	{\bibitem}
\newcommand	\s {\sigma}

\makeatletter
\newcommand\erfc{\mathop{\operator@font	erfc}\nolimits}
\makeatother
\input{epsf}
\begin{document}
\draft       

\title{Continuous phase transition in a spin-glass model without
time-reversal symmetry}
\author{G. Parisi}
\address{Dipartimento di Fisica and INFN, Universita di Roma 
{\it La Sapienza}\\
P. A. Moro 2, 00185 Roma (Italy)\\
E-Mail:  giorgio.parisi@roma1.infn.it}
\author{M. Picco}
\address{{\it LPTHE}\\
       \it  Universit\'e Pierre et Marie Curie, PARIS VI\\
       \it Universit\'e Denis Diderot, PARIS VII\\
        Boite 126, Tour 16, 1$^{\it er}$ \'etage, 4 place Jussieu\\
        F-75252 Paris CEDEX 05, FRANCE\\
E-Mail: picco@lpthe.jussieu.fr\\
        and\\
        {\it Departamento de F\'{\i}sica,\\
        \it Universidade Federal do Esp\'{\i}rito Santo\\
        Vit\'oria - ES, Brazil\\}          }
\author{F. Ritort}
\address{Departament de F\'{\i}sica Fonamental, Facultat de F\'{\i}sica\\
Universitat de Barcelona, Diagonal 647\\
08028 Barcelona (Spain).\\
E-Mail: ritort@ffn.ub.es}

\date{\today}
\maketitle

\begin{abstract}
We investigate the phase transition in a strongly disordered short-range
three-spin interaction model characterized by the absence of time reversal
symmetry in the Hamiltonian. In the mean-field limit the model is well
described by the Adam-Gibbs-DiMarzio scenario for the glass transition;
however in the short-range case this picture turns out to be modified. The
model presents a finite temperature continuous phase transition
characterized by a divergent spin-glass susceptibility and a negative
specific heat exponent.  We expect the nature of the transition in this
$3$-spin model to be the same as the transition in the Edwards-Anderson
model in a magnetic field, with the advantage that the strong crossover
effects present in the latter case are absent.
\end{abstract} 


\vskip.5pc 

\section{Introduction}
Nowadays there is large activity of research in the problem of the glass
transition from the perspective of spin-glass theory \cite{REV1,REV2}. This
interest originates from old observations by T. Kirkpatrick, D. Thirumalai
and P. Wolynes \cite{KTW} who found a striking similarity between the
dynamical equations of some mean-field spin glass models and the
mode-coupling equations for glasses. The mode-coupling equations are
characterized by the presence of a dynamical singularity at a temperature
$T_d$ below which spin-spin correlation functions do not decay to zero in
the infinite time limit signalling the breaking of ergodicity
\cite{GOTZE}. Above but close to $T_d$ the correlation functions display a
plateau which separates two different relaxational regimes (the alpha and
the beta processes). The family of models which show this behavior are
those with one-step of replica symmetry breaking (models with an infinite
number of breaking steps describe better the spin-glass behavior found in
strongly disordered magnets). These models are characterized by two
singularities or transition temperatures. One transition is purely
dynamical and corresponds to the mode-coupling transition $T_d$ previously
described.  The other transition at $T_c<T_d$ is thermodynamic and
corresponds to a temperature below which replica symmetry breaks and the
configurational entropy (also called complexity) vanishes. The transition
at $T_c$ has features of both first and second order transitions like a
discontinuity in the Edwards-Anderson order parameter and a finite jump in
the specific heat.

It is widely accepted that the origin of the dynamical transition at $T_d$
(where the relaxation time diverges and ergodicity breaks) originates from
the presence of an exponentially large number of states (exponentially
large with the system size) which trap the system for exponentially large
times forbidding the system to reach the equilibrium Gibbs measure. But it
is also clear (and this was also recognized as a strong limitation in the
original mode coupling-theory) that equilibrium below $T_d$ should be
restored in finite-dimensional systems where activated or nucleation
process (i.e. jumps over finite free energy barriers) take place in a
finite time.  What is the final theory which correctly describes the
nucleation processes taking place in glasses is one of the major open
problems. The behavior of $T_c$ in the presence of short-range interactions
is less clear. According to the mean-field picture the transition at $T_c$
(where replica symmetry breaks) could well survive in finite dimensions. At
$T_c$ the configurational entropy would still vanish. This scenario is
nothing else (but now rephrased in the spin-glass language) than the
Adam-Gibbs-DiMarzio scenario (hereafter referred as AGM) for the ideal
glass transition \cite{AGM1,AGM2}.

On top of this connection between the spin-glass theory and mode-coupling
theory, a very interesting connection has also been established between the
statics and dynamics of glassy systems in the off-equilibrium regime. In
this case, the equilibrium order parameter for spin-glasses (the so called
$P(q)$ function) is intimately related to the fluctuation-dissipation ratio
\cite{REV2}. This link between statics and dynamics, originally suggested
by the analysis of exactly solvable mean-field spin glasses \cite{CK,FM},
is actually supported by extensive numerical simulations \cite{MPRR} and
general arguments based on the assumption of linear response theory applied
to short-range spin-glass models \cite{FMPP}.

All these previous studies cannot predict how short-range corrections
change the mean-field behavior. And in particular, it is unclear how the
AGM scenario typical of a first order spin-glass transitions is modified in
finite dimensions.  The answer may crucially depends on the presence of
quenched disorder in the system and the mean-field critical behavior can be
significantly altered in finite dimensions.  Here, we will see that the
mean-field scenario \`a la AGM does not survive in finite dimensions for a
certain class of models.

A preliminary account of some parts of this work has recently appeared in
a different context \cite{MNZPPR} where a Binder-like parameter was
proposed to study replica symmetry breaking transitions.  Also, different
cases of the present model have been already studied in references
\cite{R1,R2,R3}. So our work complements these results although the case we
study here lacks time reversal symmetry sharing some features of the Ising
spin glass in a magnetic field.

The paper is divided as follows. In the section II we define the model and
the numerical algorithm is explained in section III. In section IV we
present the equilibrium results obtained by simulating small systems. In
section V we present results for the order parameter and its cumulants. In
section VI we give results for a new parameter (to be defined latter) which
unambiguously shows the existence of a phase transition. Section VII is
devoted to a discussion of the dynamical properties. Finally, in section
VIII, we summarize the main results and discuss some peculiar features of
the present model.

\section{The short-ranged p-spin model}

The present model is a short-range generalization of the multi-spin
interaction mean-field Ising spin glass defined by \cite{GM,G},

\be {\cal
H}=-\sum_{(i_1,i_2,..,i_p)}J_{i_1i_2..i_p}\s_{i_1}\s_{i_2}...\s_{i_p}\; ,
\label{eq1}
\ee 
where the spins $\s_i$, $1\le i \le N$ ($N$ is the size of the system)
can take the values $\pm 1$ and the $J_{i_1i_2..i_p}$ are quenched random
variables with zero mean and variance $p!/(2N^{p-1})$. In (\ref{eq1}) all
possible multiplets of $p$ spins interact through the random couplings
$J_{i_1i_2..i_p}$. Consequently there is no spatial dimensionality and the
model retains its full mean-field character. In order to go beyond
mean-field theory we need to suitably modify the model introducing
short-ranged interactions in a finite-dimensional lattice. A possible way
to modify eq.(\ref{eq1}) is to consider only links which couple
nearest-neighbors in a finite-dimensional regular lattice. For instance, we
could locate the spins in the vertices of the lattice and consider only a
certain set of triangles ($p=3$), a certain set of squares ($p=4$) and for
a general $p$, only a certain set of plaquettes containing $p$ spins of the
lattice \cite{MBM,CCGM}. In the presence of quenched disorder, such
constructions have been considered in several cases \cite{HR1,KRS,AFR}.  In
particular, in \cite{AFR} a simple cubic lattice with $p=4$ was
studied. Although freezing behavior was observed, no evidence for a finite
temperature transition was found in three dimensions.  Unfortunately, this
type of models with binary exchange couplings has a large ground state
degeneracy causing strong crossover effects (due to the presence of a non
trivial zero-temperature fixed point) at low temperatures.  Furthermore, in
this type of models the lower critical dimension seems to increase with
$p$. This implies that large dimensions need to be studied in order to find
a phase transition. More work is certainly necessary to identify whether
the AGM scenario is valid in finite dimensions for this type of models.

The model we are going to study here is an alternative way to include
short-range corrections. We locate $M$ different Ising spins $\s_j^{i_1
i_2...i_D}$ in each site of a regular cubic lattice. In this notation $j$
enumerates the different spins (it ranges from 1 to $M$) in a given site
with coordinates $(i_1, i_2,...,i_D)$ where $1\le i_1,i_2,...,i_D\le L$ and
$L$ is the lattice size of the cubic lattice and $D$ is the
dimensionality. The volume of the system is therefore given by $V=L^D$. The
Hamiltonian is defined as follows :

\be
{\cal H}= \sum_{1\le i_1,i_2,...,i_p\le L}\sum_{\mu=1}^{D}{\cal H}_{link}\; ,
\label{eq2}
\ee 
where ${\cal H}_{link}$ is the Hamiltonian corresponding to the link
defined by the site $(i_1,i_2,...,i_D)$ and the direction $\mu$, $1\le \mu
\le D$. In our notations, a link is a pair $(P,\mu)$ which couples the
point $P\equiv (i_1,i_2,...,i_D)$ to the nearest neighbor site in the $\mu$
direction $P+\mu\equiv (i_1,..,i_{\mu}+1,...,i_D)$. Note that in
eq.(\ref{eq2}) each link is counted only once. For each link we sum all the
possible groups of $p$ spins out of the $2M$ spins located at nearest
neighboring sites of the lattice (with $p\le 2M$).  The final expression
for ${\cal H}_{link}$ is given by

\be 
{\cal H}_{link}= -\sum_{k=1}^{p}\, \sum_{1\le a_i\le M}\,\sum_{1\le b_i\le M}
J_{link}^{(a_1...a_k,b_1...b_{p-k})}\s_{a_1}^P\s_{a_2}^P...\s_{a_k}^P\s_{b_1}^
{P+\mu}\s_{b_2}^{P+\mu}...\s_{b_{p-k}}^{P+\mu}\; .
\label{eq3}
\ee

The couplings $J_{link}^{(a_1...a_k,b_1...b_{p-k})}$ are random variables
(which take the values $\pm 1$) uncorrelated for different links $(P,\mu)$
and sets of $p$ spins $(a_1...a_k,b_1...b_{p-k})$. Other versions of the
model (for instance, $J_{link}^{(a_1...a_k,b_1...b_{p-k})} =
J^{(a_1...a_k,b_1...b_{p-k})}$, i.e. translational invariant disorder) are
also possible and they could have different properties.

As we observed in the introduction, the present model has received
considerable attention quite recently. A preliminary short account of our
work was presented in \cite{MNZPPR}. In three dimensions, an exhaustive
numerical study of the statics and the dynamics has been done in the case
$M=4,p=4$ \cite{R1} while the case $M=3,p=4$ has been studied in
\cite{R2}. So the main results on this model (except \cite{MNZPPR}) were
obtained for the $p=4$ model.  The study of the Gaussian propagators around
the mean-field limit $M\to\infty$ as well as the $1/M$ expansion were
considered in \cite{R3}. The case we study here has the crucial property
that the Hamiltonian (\ref{eq3}) does not have the symmetry under time
reversal (i.e. the global symmetry $\s_i\to-\s_i,\,\forall i$).

Here we will present exhaustive results for the case $M=2$, $p=3$ in
$D=4$. We have chosen these parameters for the following reasons:

\begin{itemize}

\item {$p=3$:} This is the simplest case which lacks time reversal
symmetry. We expect in this case more clear results about the
existence of replica symmetry breaking transitions in short-range
systems.

\item {$M=2$:} This is the simplest and non-trivial case. Larger values of $M$
require always more computational effort.

\item{$D=4$:} To be sure that we find a finite-temperature phase transition we
have studied a large dimensionality compatible with a reasonable
computational effort.
\end{itemize}

For $p=3$, $M=2$ the general Hamiltonian eq.(\ref{eq2}) reads

\bea {\cal
  H}=-\sum_{i=1}^V\sum_{\mu=1}^D(
J^{i,\mu}_{(11,10)}\s_1^i\s_2^i\s_1^{i+e_{\mu}} +
J^{i,\mu}_{(11,01)}\s_1^i\s_2^i\s_2^{i+e_{\mu}}
\nonumber\\
+ J^{i,\mu}_{(10,11)}\s_1^i\s_1^{i+e_{\mu}}\s_2^{i+e_{\mu}}
+ J^{i,\mu}_{(01,11)}\s_2^i\s_1^{i+e_{\mu}}\s_2^{i+e_{\mu}})\; ,
\label{eq4}
\eea
where the $(e_{\mu}; \mu=1,..,D)$ denote the different unit vectors in a $D$
dimensional lattice. The $J$'s are binary uncorrelated
random variables and we will consider periodic boundary conditions. 

Note that the model (\ref{eq4}) has three spin interactions. One would
simply expect the transition to belong to the class of $\phi^3$
theories. As we will see in the following sections, there is indeed a phase
transition occurring in the present model. The absence of time-reversal
symmetry in this model has crucial implications on the type of phase
transition. We anticipate that the transition is related to the breaking of
ergodicity at low temperatures, a consequence of the breaking of replica
symmetry, the crucial symmetry to describe strongly disordered systems. We
will try to clarify and give evidences on this point in forthcoming
sections. 

\section{The numerical algorithm}

We have studied the model in four dimensions using the parallel tempering
method \cite{HN,REV3}. This is a good numerical method to equilibrate
disordered systems at low temperatures. Contrarily to the simulated
tempering method, in this algorithm it is not necessary to determine the
free energy at different temperatures to reach equiprobability in the
occupancies of these temperatures.  Although the parallel tempering is a
very efficient method to surmount energy barriers it is not clear how good
the performance of the algorithm is in presence of entropy barriers
(i.e. when relaxation to equilibrium takes place along narrow channels or
gutters in phase space).

The implementation of this algorithm is quite easy. It has been widely
explained in the literature (for instance, see the reviews
\cite{REV3,BE,EM}) and we will limit ourselves to sketch the main steps of
the algorithm. We consider a set of $N_T$ copies or replicas of the same
system which stay at different temperatures ($T_i;\ i=1,..,N_T$). Each copy
or replica is then specified by a pair $({\cal C},i)$ where ${\cal C}$
denotes the microscopic configuration (i.e. the values of all the spins)
and the temperature $T_i$ of the copy $i$. We can then construct a Markov
process in the space of configurations plus temperatures which satisfies
ergodicity and detailed balance by allowing the following moves:

1) Change configuration at fixed temperature: $({\cal C},i)\to ({\cal C}',i)$
with probability

\be
P\Bigl (({\cal C},i)\to ({\cal C}',i)\Bigr )=Min(1,\exp(
-\beta({\cal H}({\cal C}')-{\cal H}({\cal C})))\; .
\label{eq5}
\ee

2) Exchange configurations of two systems at temperatures $\beta_i,\beta_j$:
$\lbrace({\cal C},i),({\cal C}',j)\rbrace\to 
\lbrace({\cal C}',i),({\cal C},j)\rbrace$ with probability

\be
P\Bigl (\lbrace({\cal C},i),({\cal C}',j)\rbrace\to 
\lbrace({\cal C}',i),({\cal C},j)\rbrace\Bigr )=Min(1,\exp(
-(\beta_i-\beta_j)({\cal H}({\cal C})-{\cal H}({\cal C}'))\; .
\label{eq6}
\ee

The first type of move is the usual change of configuration at fixed
temperature in the Monte Carlo method. The second move avoids the system to
get trapped in deep metastable minima. With this algorithm configurations
which are far from each other can be reached by allowing a single copy of
the system to extract energy from other copies through a coupling mechanism
step (2), eq. (\ref{eq6}) induced by the dynamics itself. Plainly speaking
the rest of the copies or replicas play the role of an external bath for a
given copy. If a given copy remains trapped in a deep minima of the free
energy it can escape by extracting energy from the rest of the copies. The
full Markov chain reaches thermal equilibrium when all temperatures are
equally occupied and for each temperature the conditioned probability
distribution $P({\cal C}|i)$ is a Boltzmann distribution at temperature
$1/\beta_i$.  In this way one simulates the model at different temperatures
in the same run while being always in thermal equilibrium at different
temperatures.

The time needed for the Markov process to reach the stationary distribution
(i.e. the thermalization time) depends mainly on the choice of the set of
temperatures and also on the ratio between the number of moves of the first
(\ref{eq5}) and second type (\ref{eq6}). To solve the first problem, we
have chosen a set of temperatures equally spaced in $\beta=1/T$ in such a
way that moves of the second type (\ref{eq6}) do not occur with a too small
probability at low temperatures. On the other hand, to uncorrelate the
configurations as much as possible (in order to explore maximally distant
configurations) it is convenient that the copies reach high enough
temperatures. This is accomplished by enlarging the set of temperatures in
the simulation up to twice the value of the critical temperature. For the
present model we simulated 25 temperatures ranging from $T=2.0$ up to
$T=5.0$ (as we will see this corresponds to a window of temperatures
covering the region $0.75 T_c - 2 T_c$).  As previously remarked, a good
thermalization can be achieved by choosing an appropriate number for the
ratio between the number of tempering moves of the first and second
type. If the number of moves of the second type is too small then the
system is not able to efficiently decorrelate and only a small number of
excursions into the high-temperature region are performed. In the other
extreme, the system decorrelates too fast and does not efficiently sample
the landscape at a given temperature. We have tried several schemes
intermediate between these two extremal cases and we have found that one
move of the second type for each ten moves of the first type is a good
compromise which efficiently thermalizes reasonable sizes.

We did extensive simulations for $L=3,4,5,6$ with $2^{15}=32768$ MCS
($L=3,4$) and $2^{18}=262144$ ($L=5,6$) MCS at each temperature.  For sizes
$L=3,4,5,6$ we studied $1000,600,300,100$ samples respectively. A
preliminary study for a set of 5 samples showed that this number of steps
was enough to reach thermal equilibrium for the selected sizes for the
range of temperatures studied. This preliminary study turned out to be
crucial to determine the smallest range of temperatures which allows the
system to equilibrate. If the range of temperatures selected extends down
to too low temperatures then thermalization is hardly achieved. This means
that the full Markov process associated to the parallel tempering algorithm
does not reach the stationary Boltzmann solution. Then, there is no
guarantee that thermalization is achieved neither at low nor at high
temperatures.

\section{Thermodynamic observables}

A preliminary research of the evidences for a phase transition includes the
study of the temperature behavior of extensive quantities such as the
internal energy or the specific heat. In figures 1 and 2 we show the
internal energy and the specific heat averaged over the samples for
different sizes as a function of temperature. The energy $E$ and the
specific heat $C$ were computed using the expression,

\bea E=\overline{\langle {\cal H} \rangle}\label{eq7}\\
C=\frac{\beta^2}{N}(\overline{ \langle {\cal
H}^2\rangle}-\overline{\langle {\cal H} \rangle^2})\; .
\label{eq8}
\eea

\begin{figure}
\centerline{\epsfxsize=8cm\epsffile{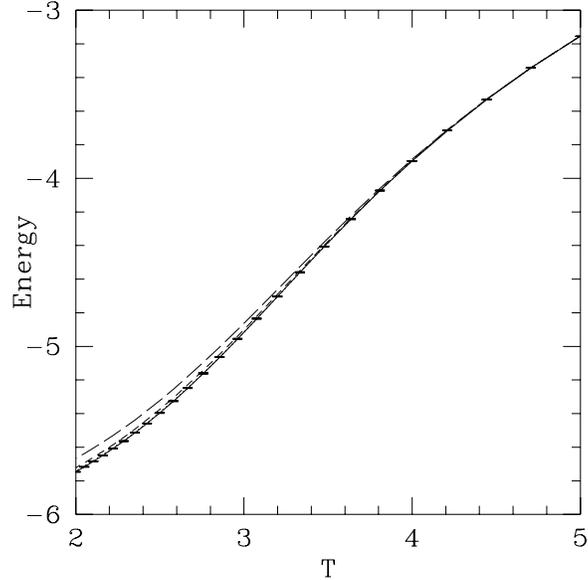}}
\caption{Energy versus temperature ($L=3,4,5,6$ correspond to long
dashed, short dashed, dot and solid lines). Error bars are shown for
$L=6$.}
\end{figure}

\begin{figure}
\centerline{\epsfxsize=8cm\epsffile{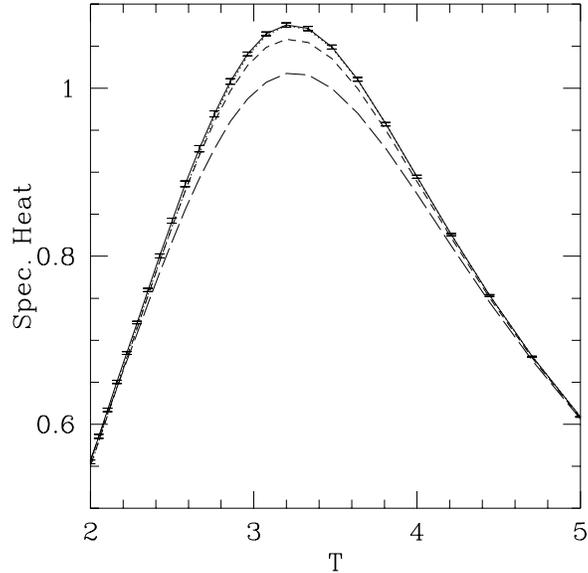}}
\caption{Specific heat versus temperature ($L=3,4,5,6$ correspond to
long dashed, short dashed, dot and solid lines). Error bars are shown for
$L=6$.}
\end{figure}

In what follows $<..>$ stands for Gibbs average and $\overline{(..)}$ for
disorder average. We note the absence of any jump in the internal energy as
well as divergence or jump of the specific heat. This is a general result
in phase transitions in strongly disordered systems and also applies in the
present model. An important feature in figure 2 (also observed in other
studies \cite{R1}) is the presence of a maximum in the specific heat at
a temperature ($\simeq 3.2$) much higher than $T_c$ (as we will see later,
$T_c\simeq 2.62$). At $T_c$ the specific heat is continuous so we expect
the specific heat exponent $\alpha$ to be negative.

\section{The order parameter}

It was shown long time ago by Edwards and Anderson that the appropriate
order parameter for spin-glasses is a measure of the temporal freezing of
the local variables (in our case, the spins) of the system. In the general
framework of spin-glass theory the order parameter is computed through the
introduction of replicas in the system. This is the natural way of
introducing the notion of a distance between two configurations in phase
space. We take two replicas of the same system $\lbrace \s_i,\tau_i\rbrace$
(i.e. two identical Hamiltonians in eq.(\ref{eq4}) with identical
realization of the couplings $J's$). Then we define the global overlap $Q$
between the two replicas $N Q=\sum_{i=1}^N\s_i\tau_i$ and evaluate its
probability distribution $P_J(q)$ averaged over the Gibbs measure
($\langle..\rangle$)

\be
P_J(q)=\langle \delta(q-Q)\rangle \; .
\label{eq9}
\ee

In our specific model the overlap is
$NQ=\sum_{i=1}^N(\s_1^i\tau_1^i+\s_2^i\tau_2^i)$ where $\s_1^i,\s_2^i$ and
$\tau_1^i,\tau_2^i$ occupy the site $i$ in the two different replicas
respectively. $P_J(q)$ gives the probability that two equilibrium
configurations $\s_i,\tau_i$ have an overlap $q$. According to the
mean-field scenario (the validity of which we would like to check for the
present model) ergodicity breaks at low temperatures and the phase space
splits up into a large number of single ergodic components or states. The
barriers separating these components diverge with the size of the system
suggesting that a symmetry is broken. This symmetry is generally referred
to as replica symmetry and it is the symmetry under the group of
permutations of a finite number of replicas. Somehow, this symmetry is
artificial (actually, it emerges from the use of the replica trick, a
general method to deal with the averaging of the logarithm of the partition
function in disordered systems). But its physical meaning is quite
appealing. Different equilibrium configurations $\s_i,\tau_i$ can take
different values according to the basins of attractions (corresponding to
different ergodic components) to which they belong. The function $P_J(q)$
is highly non trivial and this is a signature that different configurations
quite far one from each other in the phase space contribute with a finite
weight to the equilibrium partition function. Consequently, different
states have always the same free energy, internal energy and entropy per
site but they differ only in the structure of their typical
configurations. This is signalled by a non-trivial $P_J(q)$ distribution
(i.e. with several peaks at different values of $q$).

Another general consequence of the splitting up of the phase space into
different states is the possible existence of chaotic effects in the
equilibrium phase \cite{KO,RI}. A small perturbation in the Hamiltonian can
change the shape of the states as well as reshuffle their Boltzmann
weights. After perturbing the Hamiltonian the new configurations can be
very different from the initial ones. One consequence of this effect is the
existence of non self-averaging quantities. In particular, if we change
completely the microscopic realization of the disorder in the original
Hamiltonian (for instance, by changing the couplings $J$ in
eqs. (\ref{eq3},\ref{eq4}) \cite{AZFR}) the new equilibrium states differ
completely from the previous ones. In this case we do not add energy to the
system (the new and the old states have always the same energy per site)
but the reshuffling of the Boltzmann weights of the different states is
enough to change completely the form of the $P_J(q)$. Then, the $P_J(q)$ is
strongly non-self averaging, a result which has been proved in mean-field
theory and which we would like to check also in short-range systems.

The purpose of this section is to show how the study of the
$P(q)=\overline{P_J(q)}$ averaged over the disorder can yield evidence for
a phase transition in the present model. In the next section we will show
that the non self-averaging character of the $P_J(q)$ can be used as an
independent check for the transition.

A good way to characterize the $P(q)$ is through its moments. In
particular, the first moment $\overline{q}$, the second cumulant which
directly yields the spin-glass susceptibility $\chi_{SG}$, the skewness $Y$
and the Binder parameter $Z$. More precisely, if we define the average
$[f(q)]=\int dq f(q)P(q)$ (where $P(q)=\overline{P_J(q)}$) then we have :

\bea
\chi_{SG}=V [(q-[q])^2]\label{eq10}\; ,\\
Y=\frac{[(q-[q])^3]}{[(q-[q])^2]^{\frac{3}{2}}}\label{eq11}\; ,\\
Z=\frac{1}{2}(3-\frac{[(q-[q])^4]}{[(q-[q])^2]^2})\label{eq12}\; .
\eea

\begin{figure}
\centerline{\epsfxsize=8cm\epsffile{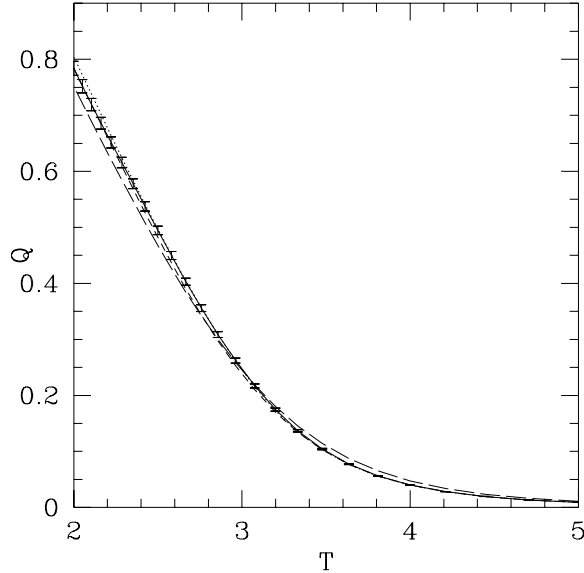}}
\caption{First moment $[q]$ versus temperature ($L=3,4,5,6$ correspond to
long dashed, short dashed, dot and solid lines). Error bars are shown for
$L=6$}
\end{figure}

In figures 3,4,6,7 we show $[q]$, $\chi_{SG}$, $Y$ and $Z$ as a function of
temperature for different sizes. In figure 3 we show the first moment as a
function of temperature. In the presence of time reversal symmetry, $[q]$
vanishes (as well as the skewness $Y$) but not in the present case. Note
that the curve of $[q]$ as a function of temperature is smooth without any
sign of a jump or discontinuity. In mean-field spin glass transitions
(continuous or discontinuous) this result is expected because the jump in
$[q]$ is proportional to $(1-m)q_{EA}$ where $q_{EA}$ is the
Edwards-Anderson parameter (the maximum value of $q$ such that $P(q)\ne 0$)
and $m$ is the replica symmetry breaking parameter (the size of the blocks
in the breaking ansatz, see \cite{BOOKS} for general introductory
textbooks). In continuous replica symmetry breaking transitions $q_{EA}$
vanishes at $T_c$ but in first-order replica symmetry breaking transitions
$q_{EA}$ is finite at $T_c$ and $m(T_c)=1$. In both cases there is no jump
in $[q]$.

\begin{figure}
\centerline{\epsfxsize=8cm\epsffile{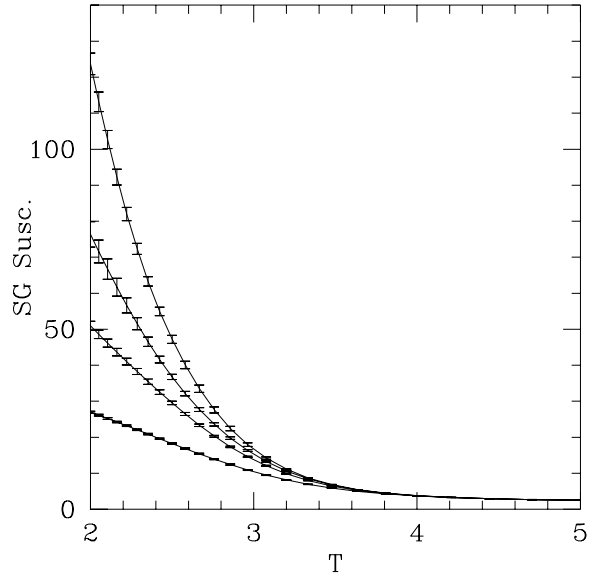}}
\caption{$\chi_{SG}$ versus temperature. From bottom to
top, $L=3,4,5,6$.}
\end{figure}

The results for the spin-glass susceptibility are more
interesting. Experimentally, spin glasses show a divergence of the
non-linear susceptibility $\chi_{nl}$ defined through the expansion

\be
M(H)=\chi_0 H+\chi_{nl} H^3 + O(H^5) \; .
\label{eq13}
\ee

It can be generally shown \cite{BOOKS} that the non-linear susceptibility
is related to the spin-glass susceptibility defined in eq.(\ref{eq10}).
Although the linear coefficient $\chi_0$ in the expansion (\ref{eq13}) does
not show any indication of $T_c$ the behavior of the non-linear term
$\chi_{nl}$ is singular at $T_c$. In mean-field spin glasses with
continuous transition $\chi_{SG}$ shows a power law divergence at
$T_c$. Contrarily, in mean-field spin glasses with discontinuous RSB
$\chi_{SG}$ shows a finite jump at $T_c$. This last feature has been
claimed to be the explanation for the violation found in experiments at the
glass transition for one of the two Ehrenfest relations \cite{THEO}. Below
$T_c$, $\chi_{SG}$ is infinite in both cases. This result is related to the
non trivial character of the $P(q)$ which has contributions from different
values of $q$. In figure 4 we show $\chi_{SG}$ for different sizes as a
function of $T$. Indeed our results in figure 4 show an algebraic
divergence of the spin-glass susceptibility $\chi_{SG}$ and a least-squares
fit of the data in the high-temperature region (where finite size effects
are negligible) yields $\chi_{SG}\sim (T-T_c)^{-\gamma}$ with $T_c\simeq
2.63$ and $\gamma\simeq 1.0$. A finite size scaling plot of the data for
$\chi_{SG}/L^{2-\eta}\sim (T-T_c)L^{1/\nu}$ is shown in figure 5 with
$\nu\simeq $2/3$, \eta\simeq 1/2$.  This is in agreement with the exponents
relation $\gamma=(2-\eta)\nu$ and with the previously estimated value of
$\gamma$. Using the hyperscaling relation $\alpha=2-D\nu$ we get
$\alpha\simeq -2/3$ in agreement with the absence of any singularity or
jump in the specific heat (see figure 2).  An independent measure of $\nu$
will be obtained in the following section. Anyway, from this first data, we
may conclude that the divergence of $\chi_{SG}$ is related to the
divergence of a correlation length at $T_c$.

\begin{figure}
\centerline{\epsfxsize=8cm\epsffile{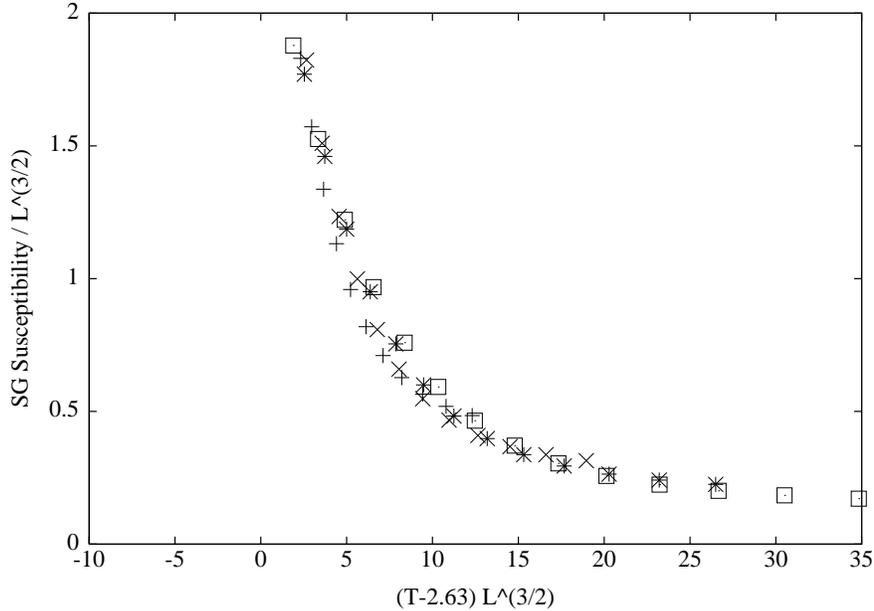}}
\caption{Finite size scaling of the spin-glass susceptibility. Our data
is compatible with $\eta\simeq 1/2, \nu=2/3$.} 
\end{figure}

\begin{figure}
\centerline{\epsfxsize=8cm\epsffile{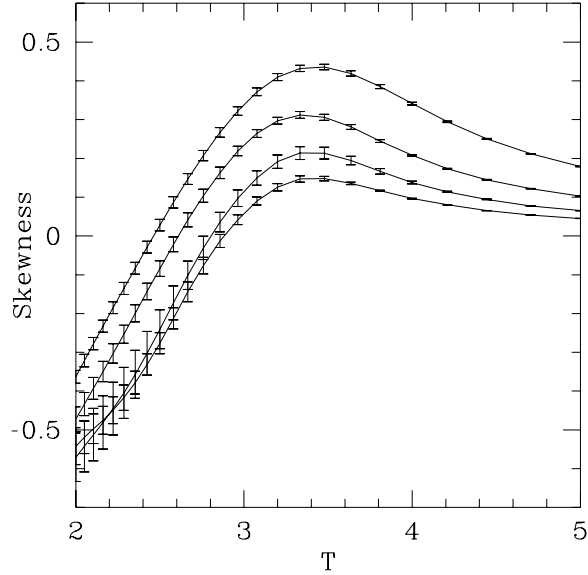}}
\caption{Skewness versus temperature. From top to
bottom, $L=3,4,5,6$.} 
\end{figure}

\begin{figure}
\centerline{\epsfxsize=8cm\epsffile{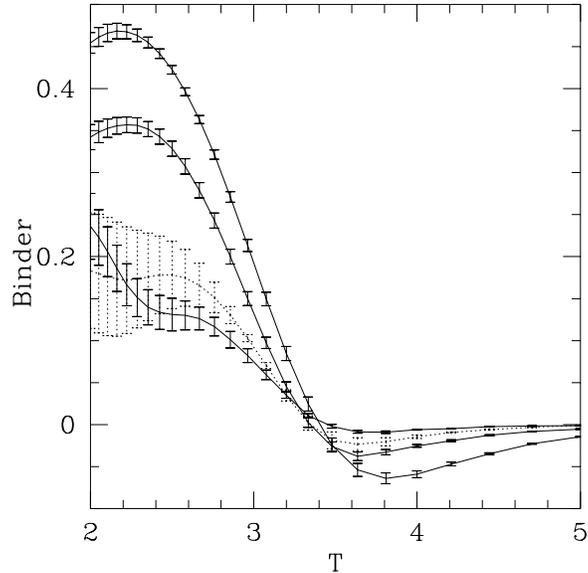}}
\caption{Binder parameter versus temperature. From bottom to top in the
high temperature phase, $L=3,4,5,6$.}
\end{figure}

Figures 4 and 5 are the first evidence for a phase transition in the
model. Figures 6 and 7 show the skewness and the Binder parameter as a
function of temperature. Because of the adimensional character of these
quantities one expects that they should be universal in the critical point
and related to the amplitudes of the renormalization group flow
equations. In the large volume asymptotic regime the value of the skewness
$Y$ and the Binder parameter $Z$ should be volume independent at
$T_c$. Consequently $T_c$ should manifest as a common crossing point of the
curves corresponding to different system sizes. A common crossing point can
be hardly identified in figures 6 and 7.  Nevertheless, the fact that $Y$
and $Z$ do not vanish at low temperatures is a sign for a non trivial low
temperature phase.  Actually these two figures yield few information about
the transition and it is hard to guess what is the character of the phase
transition.  Let us note that strikingly similar results were obtained for
the Ising spin glass in a magnetic field \cite{PR}.

\begin{figure}
\centerline{\epsfxsize=8cm\epsffile{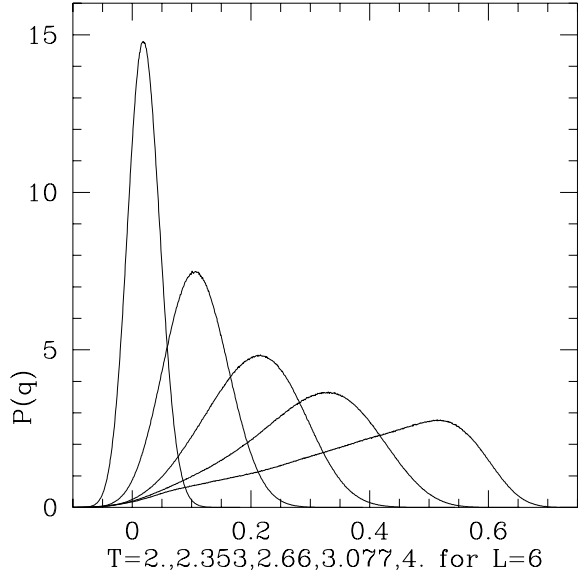}}
\caption{$P(q)$ for $L=6$ at different temperatures (from left to right,
$T=4,3.077,2.66,2.353,2.0$).}
\end{figure}

Contrarily to what is found in models with time-reversal symmetry the
$P(q)$ is not symmetric around $q=0$. This explains why $[q]$ and the
skewness ($Y$) are non zero. Moreover, the fact that $Y<0$ is related to
the asymmetric character of the $P(q)$. This is clearly shown in figures
8,9 and 10. Figure 8 shows the $P(q)$ for the largest simulated size $L=6$
at different temperatures $T=4.0,3.077,2.66,2.353,2.0$. Note that above
$T_c$ the $P(q)$ is Gaussian and develops a non-trivial shape at low
temperatures with a non vanishing tail which extends down to values of $q$
close to zero (the presence of this tail is clearly appreciated plotting
the vertical axis in a logarithmic scale).

Except for the fact that the $P(q)$ is not symmetric around $q=0$ this
behavior is reminiscent of what is observed in \cite{R1} for the $p=4$
case. Figures 9 and 10 show the $P(q)$ for different sizes at $T=2.66$ and
at $T=2.0$ respectively.

\begin{figure}
\centerline{\epsfxsize=8cm\epsffile{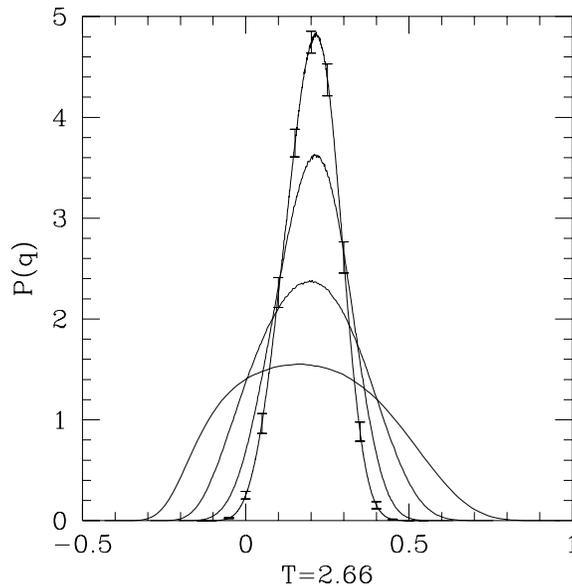}}
\caption{$P(q)$ at $T\simeq T_c$ for $L=3,4,5,6$. For the sake 
of clarity, we show the error bars only for $L=6$ and only for some values
of $q$.} 
\end{figure}

\begin{figure}
\centerline{\epsfxsize=8cm\epsffile{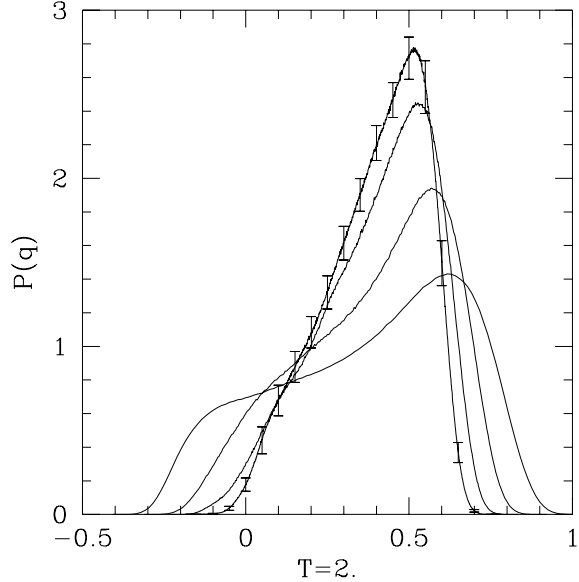}}
\caption{$P(q)$ at $T=2.0$ for $L=3,4,5,6$. Errors are shown for 
$L=6$, for some vales of $q$.}
\end{figure}

\section{A new evidence for the transition}

A new evidence for the existence of a phase transition can be obtained from
the study of the sample to sample fluctuations of the order parameter
function $P_J(q)$. Our main observation originates from recent results
obtained by F. Guerra \cite{GUERRA}. Guerra has shown that sample to sample
fluctuations of the cumulants of the order parameter distribution $P_J(q)$
are Gaussian distributed in the thermodynamic limit. Let us now define a
sample dependent (i.e. $J$ dependent) susceptibility through

\be
\chi_{SG}^J=V(\langle q^2\rangle-\langle q\rangle^2)\; .
\label{eq14}
\ee

Note that $\overline{\chi_{SG}^J}$ is different from $\chi_{SG}$ defined in
eq.(\ref{eq10}). It can be proved \cite{GUERRA} that the following
relationship is fulfilled in spin glasses below $T_c$

\be
G=\frac{\overline{(\chi_{SG}^J)^2}-\bigl (\overline{\chi_{SG}^J}\bigr)^2}
{\overline{V^2\langle(q-\langle q\rangle)^4\rangle}-
\bigl(\overline{\chi_{SG}^J}\bigr)^2}=\frac{1}{3}\; ,
\label{eq15}
\ee
where, as before, $\overline{(.)}$ means average over the quenched
disorder.

The interest of defining the parameter $G$ is that it vanishes above the
transition temperature in the disordered phase where sample to sample
fluctuations of the $P_J(q)$ disappear in the $V\to\infty$ limit.  Similar
information to that obtained from (\ref{eq15}) can be also gathered from
the sample to sample fluctuations of $\chi_{SG}^J$,

\be
A=\frac{\overline{(\chi_{SG}^J)^2}-\bigl(\overline{\chi_{SG}^J}\bigr)^2}
{\bigl(\overline{\chi_{SG}^J}\bigr )^2}\; .
\label{eq16}
\ee

In principle, eq. (\ref{eq16}) yields also non trivial behavior in the low
temperature phase even though (in contrast to $G$) it does not necessarily
converge (in the thermodynamic limit) to a temperature independent
value. Both parameters ($A$ and $G$) are good indicators of the transition
although only $A$ gives a precise answer to the question whether the order
parameter is self-averaging or not. The reason \cite{COMMENT} is that $G$
may be finite even when the numerator and denominator in eq.(\ref{eq15})
vanish. Actually $A$ is the numerator of $G$ so it gives a precise
information whether self-averaging is satisfied \cite{RESPONSE}.

Note that $G$ is a parameter which plays the same role as the usual Binder
parameter $g$ in ferromagnets and is given (in the $V\to\infty$ limit) by
$G(T)=(1/3)(1-\Theta_H(T-T_c))$ where $\Theta_H$ is the Heaviside theta
function. In RSB transitions (\ref{eq15}) goes to zero (as the size $V$
increases) as $1/V$ for $T>T_c$ but converges to a finite value for
$T<T_c$. We expect the critical temperature (where RS breaks) to be
signaled by the crossing of the different curves corresponding to different
lattice sizes.

\begin{figure}
\centerline{\epsfxsize=8cm\epsffile{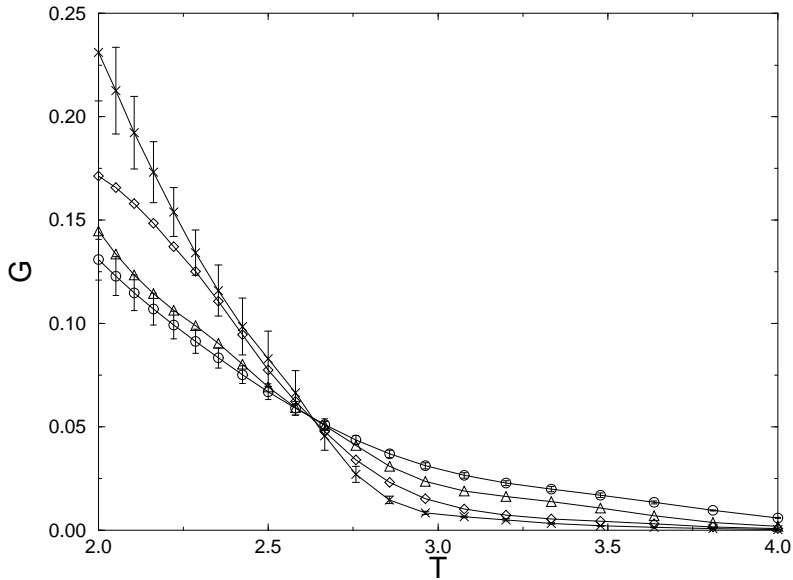}}
\caption{Parameter $G$ for $L=3,4,5,6$ (open circles, triangles,
diamonds and crosses respectively).}
\end{figure}

Our results for $G$ and $A$ are shown in figures 11 and 12.  Both figures
show essentially the same result, i.e. the curves for $G$ and $A$ for
different sizes display a common crossing point located approximately at
$T_c\simeq 2.63$ in agreement with the result derived in the previous
section from the divergence of the spin-glass susceptibility. Assuming for
the parameter $G$ the following scaling behavior $G(T)=\hat{G}(L/\xi)$ with
$\xi\sim (T-T_c)^{-\nu}$ then $(dG/dT)_{T=T_c}\sim L^{1/\nu}$. In figure 13
we show the scaling behavior for $A$ and $G$.  The scaling plots for $A$
and $G$ yield a more precise fit to the critical exponent $\nu$ because it
does involve only one free parameter ($T_c$ was obtained looking at the
crossing of the different curves).  A good estimate for $\nu$ yields
$\nu\simeq 2/3$ for both $A$ and $G$ but precision is not good enough to
exclude a slightly smaller value (such as $\nu=1/2$). The value $2/3$ is in
good agreement with the one obtained from the divergence of the spin-glass
susceptibility.

\begin{figure}
\centerline{\epsfxsize=8cm\epsffile{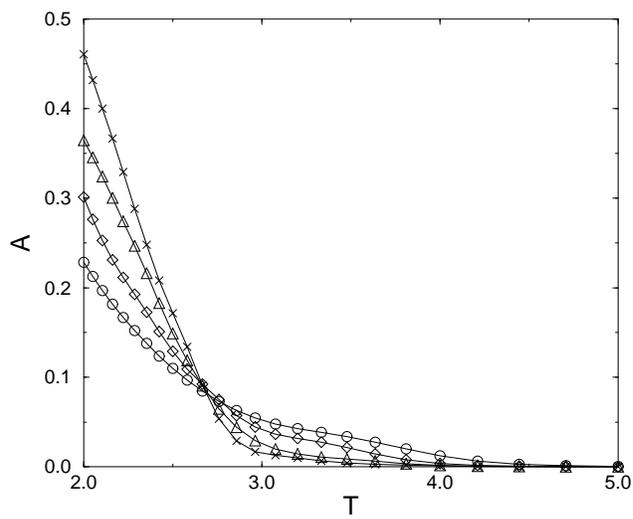}}
\caption{Parameter $A$ for $L=3,4,5,6$ (open circles, rhombi, triangles
and crosses respectively)}
\end{figure}

\begin{figure}
\centerline{\epsfxsize=8cm\epsffile{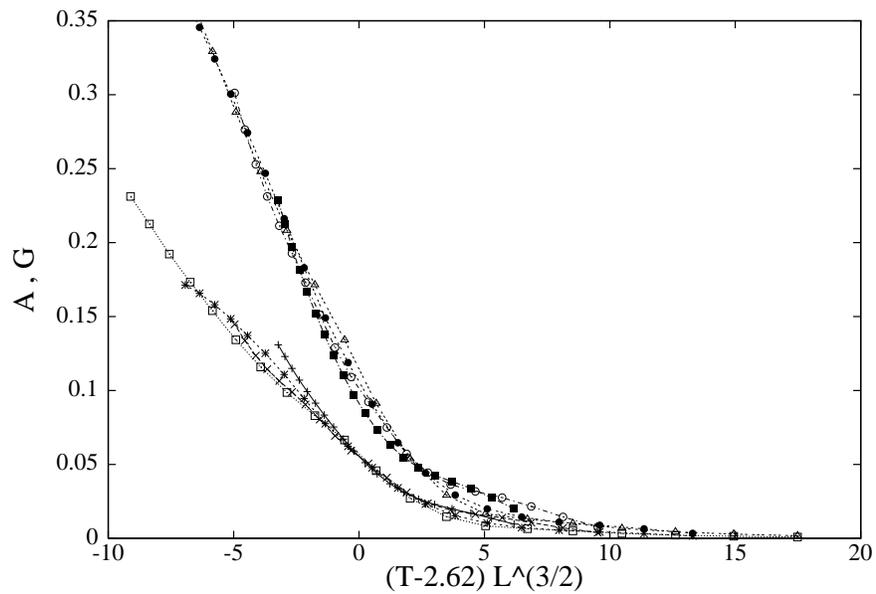}}
\caption{Parameters $A,G$ for different sizes $L=3,4,5,6$ 
versus $(T-2.62) L^{3/2}$}
\end{figure}

Figures 11 and 12 clearly suggest that replica symmetry breaking takes
place below $T_c$. This supports the result that the breaking of ergodicity
is intimately related to the non self-averaging character of the order
parameter. This result is in contradiction with heuristic arguments by
Newman and Stein \cite{NS} who have suggested that self-averaging should be
automatically restored in short-range systems due to the translational
invariance symmetry of the lattice.

\section{The dynamical exponent z}

Now that we have corroborated the existence of a thermodynamic phase
transition in the model we would like to learn more about its nature. In
particular we would like to clarify whether the relaxation time (which is
the analogous of the shear viscosity of real glasses) shows an anomalous
behavior in the vicinity of the glass region. It is well known that an
activated behavior in the relaxation time is one of the main
characteristics in real glasses. On the other hand, a power law divergence
of the relaxation time is a signature for a second order phase transition
where a massive mode vanishes.

We have computed the equilibrium time correlation function $C(t)$ at
several temperatures above the estimated $T_c$.  $C(t)$ is defined through

\be
C(t)=\frac{1}{V} \sum_{i=1}^V(\s_1^i(0)\s_1^i(t)+\s_2^i(0)\s_2^i(t))\; .
\label{eq17}
\ee

Compared to similar studies undertaken in the $p=4$ case \cite{R1,R2}, the
analysis in the present case turns out to be more difficult.  This is due
to the fact that in the absence of time reversal symmetry in the
Hamiltonian, $C(t)$ does not decay to zero and there is one more unknown
parameter ($C(\infty)$). To have a reasonable estimate of $z$ we did two
types of measures. On the one hand we measured $C(t)$ and fitted it to a
stretched exponential form

\be
C(t)=q+(2-q)\exp(-(t/\tau)^{\beta}) \; ,
\label{eq18}
\ee 
with three free parameters $q$ ($C(\infty)$), $\tau$ (the relaxation time)
and $\beta$ (the stretching exponent). Note that $C(t)$ is normalized such
that $C(0)=2$.  Figure 14 shows some of the fits which turn out to be quite
good. In this way we got some estimates for $\tau$ which unfortunately are
not very precise to determine the value of $z$.  To be more precise one
should include a power law term $t^{-\alpha}$ multiplying the exponential
term in the fitting function (\ref{eq18}) as was done by Ogielsky in the
study of the three dimensional Edwards-Anderson model \cite{OGI}.
Unfortunately including a term of this type in (\ref{eq18}) introduces too
many free parameters into $C(t)$ making fits poorly predictive.
Nevertheless, from figure 14, we may conclude that relaxation turns out to
be very slow close to $T_c$

Our values estimated for the relaxation time exclude any activated
behavior. This excludes the existence of a viscosity anomaly as well as the
existence of two step relaxation processes in this model. The same
conclusion was reached in the $p=4,D=3$ case by studying the $C(t)$
\cite{R1}.

An estimate for $z$ can be obtained by studying the off-equilibrium decay
of the order parameter \cite{BHB} or the internal energy \cite{CP}. This
last case has been applied also to the study of structural glass models
\cite{PA} as well as in the $p=4$ case \cite{R1}. In this case, one studies
the decay of the internal energy starting from a random initial
configuration at $T=T_c$ and using a fit to a power law behavior of the
following form

\be
E(t)=E(\infty)+At^{-\lambda}\; .
\label{eq19}
\ee

Note that this is an off-equilibrium measure which is expected to yield the
equilibrium dynamical exponent. Under the assumption that hyperscaling is
valid, and using simple scaling relations one obtains the exponents
relation $\lambda=(d-1/\nu)/z$. In figure 15 we show the decay of the
internal energy at the estimated $T_c$. A good fit is obtained with
$\lambda=0.55\pm 0.1$ which yields $z\simeq 4.5\pm 1$.  This value for $z$
is very similar to the one found in the Edwards-Anderson model in four
dimensions. Still, in this model the divergence of the relaxation time is
not fast ($z\nu \simeq 3$) if compared with that of the the $M=3,p=4$ model
\cite{R1,R2} ($z\simeq 7$, $z\nu\simeq 6\pm 1$). The value of $z\nu$ being
not very large in our model (at least, compared to those generally found in
spin-glass models in three dimensions) explains why we succeeded in getting
very clean results through finite-size scaling for the existence of a phase
transition. Thermalization was easier to achieve.

\begin{figure}
\centerline{\epsfxsize=8cm\epsffile{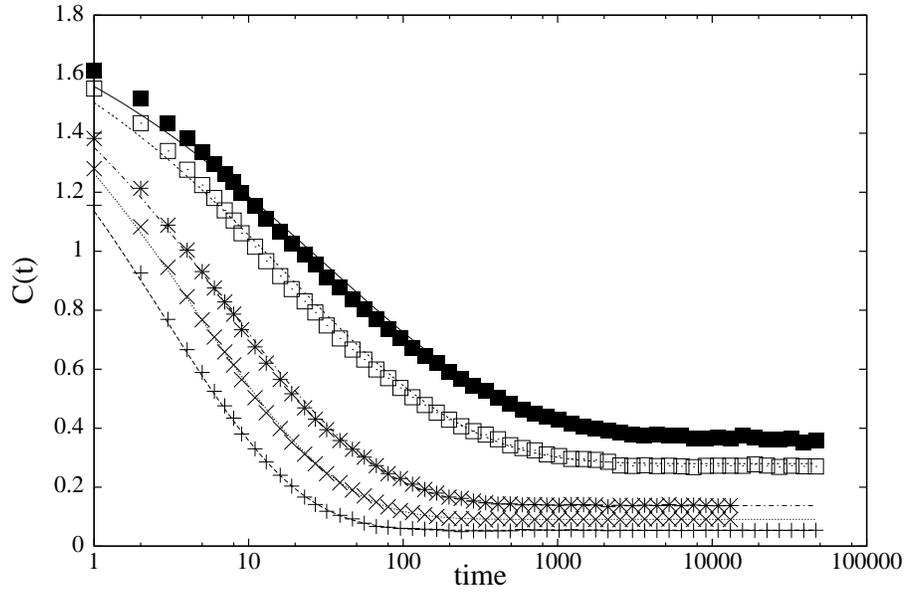}}
\caption{$C(t)$ at $\beta=0.26$ (+), $\beta=0.28$ (x), $\beta=0.30$ (*),
$\beta=0.34$ (empty squares), $\beta=0.36$ (filled squares) for one
sample with $L=20$}
\end{figure}

\begin{figure}
\centerline{\epsfxsize=8cm\epsffile{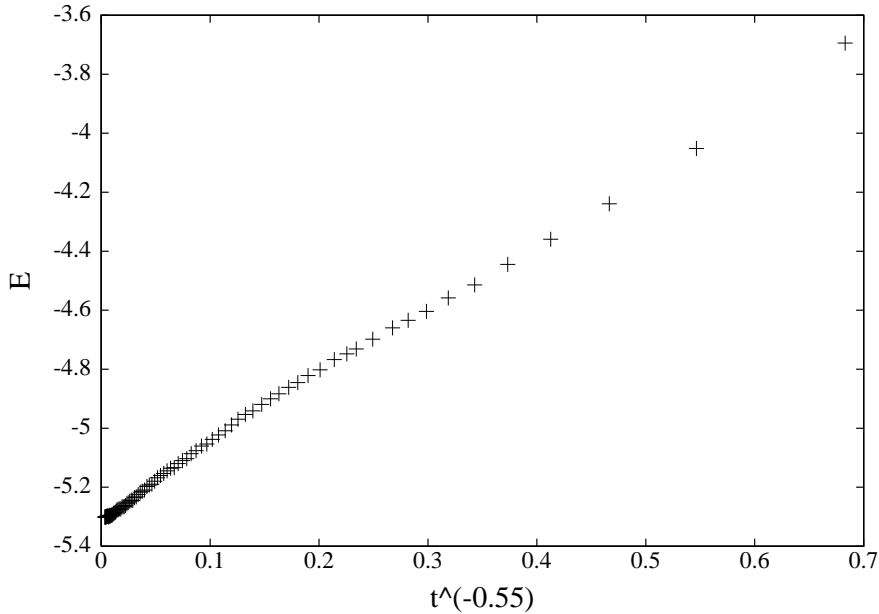}}
\caption{Energy versus $t^{-0.55}$ at $T=2.62$ for one sample with $L=20$}
\end{figure}

\section{Discussion}

In this paper we have investigated the critical behavior of a three spin
model in finite dimensions. The motivation was to understand how
short-range effects modify the AGM mean-field scenario for the spin-glass
transition. Moreover the Hamiltonian for the present model $(M=2,p=3$) has
no time-reversal symmetry. In the presence of a thermodynamic transition
this lack of time reversal symmetry has far reaching consequences on the
nature of the transition.

We have established (through finite size scaling methods) the existence of
a phase transition without latent heat and with an algebraic divergence of
the spin-glass susceptibility. On the other hand, we find indications that
the relaxation time diverges according to an algebraic power law as in
ordinary continuous phase transitions excluding the presence of an
activated relaxation time.

Consequently, we are lead to the conclusion that the first order character
of the transition present in the mean-field limit ($M\to\infty$) is lost in
finite dimensions. In particular, our results $\nu\simeq 2/3$, $\eta\simeq
1/2$, $\gamma\simeq 1$, $z\simeq 4.5$ yield reasonable fits to all the
data. Moreover, these exponents yield a negative value for the specific
heat exponent in agreement with the fact that there is not jump or
divergence in the specific heat at $T_c$. Note that $\alpha$ vanishes in
mean-field so the main effect of finite-dimensional corrections is to
decrease the value of $\alpha$.  A negative value for $\alpha$ was
apparently also obtained for $p=4$ in three dimensions for
$M=3,4$. Although large size simulations in $p=4,D=3$ \cite{R2} yield a
smaller value of $\nu$ (compatible with $\nu=\frac{2}{D}$ and hence
$\alpha=0$) we must exclude this possibility from the absence of any jump
in the specific heat at $T_c$.

Quite long ago Gross, Kanter and Sompolinsky, from the solution of the
mean-field Potts glass \cite{GKS}, suggested the possibility that $\nu=2/D$
could be valid in finite dimensions similarly as happens for pure systems
(with the corresponding relation $\nu=1/D$). In particular, the explanation
for the rounding of the phase transition due to finite size effects would
be very similar to the explanation valid in first order transitions in pure
systems \cite{NN,FB} but now modified to account for the presence of
randomness. If the transition in finite dimensions were first order then we
would expect the validity of the relation $\nu=2/D$ as well as the absence
of upper critical dimension. Our numerical results tend to discard this
possibility.

Note that the model we are considering here has no time-reversal
symmetry. Consequently, any thermodynamic transition cannot be associated
to the breaking of an original symmetry of the Hamiltonian. In this
respect, the transition we are facing reminds a lot of the mean-field
transition of spin-glasses in a magnetic field.  We believe that the type
of transition presented here is one of the most clear examples of second
order phase transition in strongly disordered systems where replica
symmetry breaks. Figures 11, 12 and 13 offer good evidence for this result.

Finally we would like to comment on the behavior of the entropy of the
model as a function of the temperature. One prominent prediction in the AGM
scenario is the collapse of the configurational entropy at $T_c$.
Obviously in the present model we do not expect that the configurational
entropy vanishes at $T_c$ since the transition is continuous. In figure 16
we plot the total entropy (which is the sum of the configurational and its
intrastate part) as a function of $\beta$ obtained numerically by
integrating the internal energy between $\beta=0$ and $\beta$. As $\beta$
increases (data is shown between $\beta=0$ and $\beta=0.32$) the entropy
becomes steadily linear and seems to extrapolate to zero at a finite value.
A linear fit to the last set of data yields $S(\beta)\simeq 5.78
(0.364-\beta)$ which vanishes at $T_c\simeq 2.75$, a result strikingly
close to the previously estimated value of $T_c$ through finite-size
scaling methods. This value of $T_c$ is quite stable against data inclusion
or exclusion in the fit. Actually the same linear behavior is observed for
the entropy when plotted versus the temperature. A linear fit yields in
that case, $S(T)\simeq 0.536(T-2.64)$ a result still very close to previous
estimates of $T_c$. Consequently the total entropy (and also the
configurational entropy) seems indeed to vanish at a temperature slightly
above $T_c$. This is non sense because above $T_c$ the entropy must always
be finite. We conclude that the entropy must stop decreasing at
temperatures close to $T_c$ and depart from the linear behavior again.
Actually, this is what we expect from the presence of the maximum of the
specific heat (figure 2). Using the relation $C=T\frac{\partial S}{\partial
T}$ we safely predict a breakdown of the linear behavior at $T\simeq 3.2$
($\beta\simeq 0.31$) where the entropy should start to form a plateau.

To conclude, we have found that short-range effects in some class of models
(with non-translational invariant disorder) make the transition to become
second order. The type of transition should be the same to the one expected
for the Edwards-Anderson model in a magnetic field. The main difference is
that, in the last case, the parameter space contains the temperature and
the field while in the former case the only parameter which controls the
phase transition is the temperature. This implies strong crossover effects
in the critical region for the Edwards-Anderson model in a field due to the
proximity of the zero-field fixed point. Such crossover effects are not
present in the present model making the determination of the critical
behavior much simpler.  It would be very interesting to extend the research
of strongly disordered spin models without time-reversal symmetry to other
cases, such as models with translationally spatial invariant disorder, to
understand under which conditions the first order character of the
mean-field transition survives in finite dimensions.

\begin{figure}
\centerline{\epsfxsize=8cm\epsffile{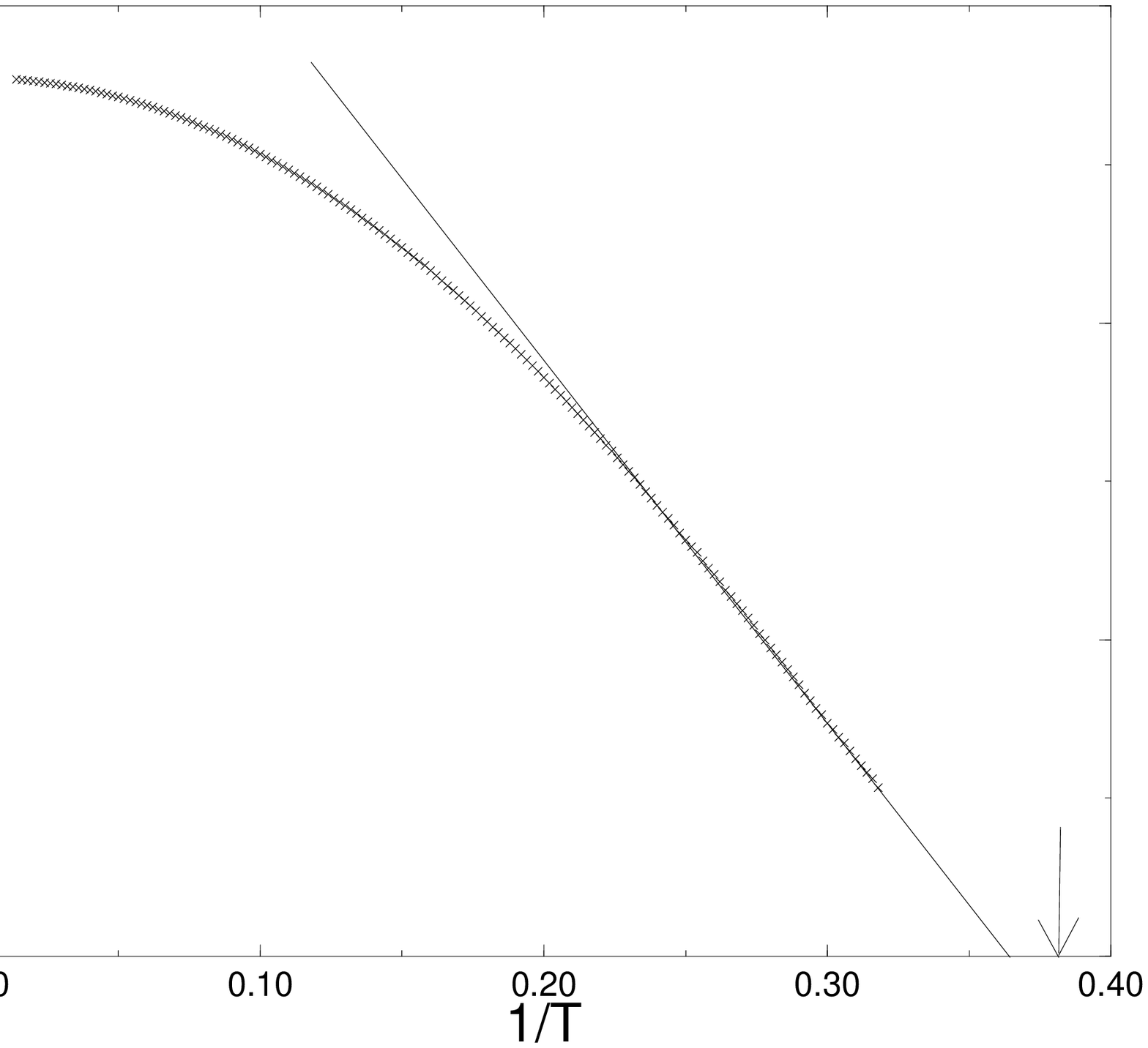}}
\caption{Entropy of the model versus temperature. The continuous line
is a linear extrapolation of the last set of points. Data were
obtained simulating a sample with $L=10$. The arrow corresponds to the
estimated value of $T_c$.}
\end{figure}

{\bf Acknowledgments}. We acknowledge Matteo Campellone for
discussions and a careful reading of the manuscript. F.R. acknowledges
FOM and the University of Amsterdam for financial support where part
of this work was done and Ministerio de Educacion y Ciencia in Spain
for support through project PB97-0971. M.P. acknowledges CAPES
(Brazil) for financial support and the Physics Department of the UFES,
and in particular J.~Fabris, for its hospitality.

\hspace{-2cm}

\vfill
\end{document}